\documentstyle[12pt,epsf]{article}
\catcode`\@=11 \@addtoreset{equation}{section}  %%
\renewcommand{\theequation}                     %%
         {\arabic{section}.\arabic{equation}}   %%
\title{Super-Relativity and State-Dependent Gauge Fields}
\author{P. Leifer}
\date{ODIN Technologies, Ltd., Yokneam, Israel} 
\begin{document}
\maketitle
\begin{abstract}
Objective interpretation of the quantum states leads
to the enhancement of the principle  of `relativity to the measuring device' to the principle of `relativity to embedding into environmental field' and further - to the super-relativity.
In fact the geometric phase works rely on tacitly thinking version of such principle.
The geometric phase approach refers to the 
symplectic invariants of the state space
(or parameter space) independent on dynamical 
(in the space-time) behavior of ``macroscopic'' 
quantum systems. As far as understood the geometric 
phase concept is applicable to the really
``microscopic'' quantum systems like 
`elementary particles', due to a generality, as well.
However unification of the fundamental 
interactions including gravity requires the geometric 
(in the state space) description of quantum dynamical  
variables. Therefore the state space
geometry should be non-distinguishable from the 
quantum dynamics. The new kinds of dynamical 
invariants (quantum integrals) associated with symmetric 
part of K\"ahler structure arise in such description.
These integrals are relating to some analog of non-Abelian 
Yang-Mills unified fields.

\end{abstract}
\vskip .2cm
\section{Introduction}
The eyewitness role is seldom achievable for a
detective. Hence, only some traces of a crime 
are really observable. Therefore it would be 
very strange to demand from the detective to 
operate with observable facts only during the 
crime investigation. He should, of course,
use some evidences and facts, but the most
important to be able to connect these fragments 
into the entire `real' picture of the crime. 
Physicist frequently acts like the detective trying 
to find the rational explanation of an observable 
phenomenon. That is why Einstein emphasized that 
it would be incorrect to base physical theory on 
the observable quantities only.
Therefore, the observability principle, from this 
point of view, should give up 
one's seat to the objective principle of the relativity 
to the embedding into environmental field as an 
enhancement of 
the `relativity to the measuring device' principle. 
`Principle relativity to the measuring device' was 
declared  by Fock in quantum theory as the 
generalization of relativity to the reference frame 
in classical physics \cite{Fock1}. We will avoid here the criticism of Fock's philosophy which was in the 
contradiction with (correct from our point of view) Dirac's objective interpretation of quantum state of a system \cite{Dirac1}.  Such comprehension
of the quantum state leads to the consistent 
development of Fock idea of the `relativity to the 
measuring device';- namely,
{\it relativity to embedding into environmental field}. 

During last years I have understood that 
super-relativity
principle \cite{Le1,Le2,Le3,Le4} may be 
understood as the principle of relativity to 
embedding into environmental 
field in a `strong' interpretation.
The `weak' interpretation is simply trivial
statement, since our scientific paradigm based
on the assumption that different physical devices
(let's say accelerators in Protvino and in the CERN, 
i.e. different
background fields) give us qualitatively identical
results on proton's collisions. But this  
form of the invariance principle
is very restrictive since there is the energetic 
threshold for external field beyond that we have new 
generations of the leptons, etc. In order to avoid 
this limitation, one should admit the 
particle identification as a function
of the measuring equipment (background field): 
{\it different choice
of the measuring (generating) device (i.e. `external 
fields') may lead to the different
combination of internal degrees of freedom, i.e.
to the quantum particle mutation}.
It is clear that the `strong' invariance principle
may be realized on very deep level. Presumably
it may be Plank's scale level. Realistic physical
picture on this level is absent, but at least in one
thing we can be sure that the electrodynamics itself 
is not acceptable on this level as a paragon gauge 
theory. 
Namely, pointwise charged particles is very deep
defect of the theory capable to spoil any efforts 
in mimic generalization of the gauge theory 
\cite{Jackiw}. 
However, extended (in the space-time)
quantum particles are localizable in {\it a 
functional
state space}. Therefore in the state space 
we have to have a state-dependent
gauge principle \cite{Le1}. Herein more 
general kind of
gauge invariance has been proposed than one 
mentioned in Jones'es work \cite{Jones}. 
It will applied to the fundamental fields (FF) 
defined 
on projective Hilbert space $CP(N-1)$ \cite{Le4}. In
principle, one can use $CP(\infty)$ but frequently
finite dimension projective space is quite enough
for practical targets. 

Some aspects of $CP(N-1)$ using
as a phase space discussed already widely, however,
for different physical aims
(see for example well known articles in this 
direction \cite{PV,Heslot,CMP,ZF}).
These approaches should be 
changed for our targets as follows:

1. The ordinary quantum mechanics is obviously 
linear
with very high degree of accuracy. It is clear that the foundations 
of quantum mechanics are rooted into the nonlinear 
quantum field theory of the `elementary' particles.
Therefore the linear character of quantum mechanics
is an approximation of the more deep theory but of course
is not a particular case of classical
physics (even advanced one) \cite{Heslot}. 
We propose $CP(N-1)$ 
as an underlying unified physical construction for 
the field foundation of quantum theory. Namely, the
points of projective Hilbert space represent 
states of all
conceivable background fields (matter) created 
by self-consistent
Universe cosmic potential (see below). From the 
mathematical
point of view projective Hilbert space is the 
base of tangent fiber bundle. In each tangent
space when local coordinates are close to 
zero, the
curvature is negligible and one has the 
linear picture
of the ordinary quantum mechanics (the 
tangent space
approximation).   

2. Geometric description of the gravity is 
3. achievable 
due to the fact of the independence of accelerations 
of freely falling masses. The space-time
behavior of quantum particles is definitely different.
Hence, a different criterion than accelerations 
equivalence,
and different arena rather the space-time should 
be used for the geometric unification of the 
fundamental 
interactions in the quantum case. First of all 
the notion
of a quantum state should be used instead
of material point notion. Then, I assume that 
desirable unification of all fundamental
interactions may be achieved if we take into account
the universal projective Hilbert space geometry of 
quantum states. In particular, the
problem of the comparison of two quantum states
in the different background fields leads to the 
`parallel transport' in some generalized sense 
and newly understood affine connection.  
In fact $CP(N-1)$ geometry is the unique common property of
all known quantum particles since their rays of states
obey to identical geometric laws  under unitary dynamics.
Local coordinates in the projective Hilbert space
will correspond boson fundamental field (FF) \cite{Le4}.
It looks like the phenomenological Lagrangians method
but boson fields presented by the points
of the $CP(N-1)$, are now coordinates of dynamical 
field variables like $D^i_\sigma$ (see below) realizing
the nonlinear representation $SU(N)$ group.

3. Riemannian structure treated in the works 
 \cite{PV,Heslot,CMP} as an evidence of the
statistical character theory.  We develop the 
deterministic
foundation of quantum theory assuming that 
statistical
properties arise under space-time embedding of 
underlying 
nonlinear (K\"ahlerian) pre-dynamics.

4. Due to the positive curvature of the
projective Hilbert state space these ``geometric
bosons'' are effectively interacting and 
being condensed (like the spin complex) obey
nonlinear differential equations of motion.
Presumably one can find some Skyrmion-like (soliton)
solution of these equations determined by the 
projective structure of $CP(N-1)$. Further they may 
be quantized like boson or fermion \cite{Aitch}. 

5. I will discuss here the vacuum 
excitations modes of the ``fundamental field'' 
\cite{Le4}. We assume that the FF may be identified
with the field of a self-consistent global Universe 
potential $U_{global}$ and,
hence, may treated as a vacuum for each concrete 
quantum system. The concrete quantum system
will be defined in tangent space $T_{\pi}CP(N-1)$
at the origin point $\pi = (\pi^1,...,\pi^{N-1})$
by the set of dynamical variables
(generators of the $SU(N)$).   

6.  Poincar\'e group arises as covering group
for ``question'' dynamical variables (logical spin 1/2)
treated as a local space-time coordinates.

7. The area of pulse (with the action dimension) 
transmitted to the quantum system will be treated 
herein as the invariant measure 
of the fundamental interaction. 
In order to demolish the ambiguous of the external field 
one should demolish the difference between environmental 
field and `particles'.  In fact quantum field theory
made already important steps in that direction.
Namely, all fields should be `secondly quantized'
i.e. represented as a result of the action by specified
creation operators on the standard vacuum state $|V>$.
Procedure of the second quantization has been proposed by
Dirac. This consist of the prescription that the Fourier components of the wave function (c-numbers) should be 
changed by operators (q-numbers) with definite 
commutation relations.
However there are a lot of problems with such
approach, namely: all operators corresponding to 
particles
one should be introduced `by hand'; commutation 
relations between these operators
should be given a priori (boson or fermion);
such commutation relations lead to highly singular
functions and problematic description suffering 
on UV divergences.  
Therefore some different version of the `second
quantization' as unified quantum dynamics we must
build. I will establish in the present work
only general framework
for such kind of quantization.  
\vskip .2cm
\section{Vacuum and Self-Consistent Potential of Universe}
\vskip .2cm
Nobody knows the real self-consistent Universe potential
$U_{global}$ and corresponding vacuum state $|V>$. However 
we have some hints appeared from the foundation of 
general and special relativity.

It is well known that 
postulate of light's velocity invariance in special 
relativity, was rejected by Einstein during development 
of the general relativity. Namely, light's velocity in
the general relativity is a function of the gravitation
potential. Since all material objects have energy (mass)
they produce local in Universe perturbation of the
global potential and, hence, light's velocity.

I will assume there is the global (gravitation or, 
probably,- universal?) self-consistent potential of 
the Universe $U_{global}$ with the value $c^2$. 
It is clear that now $E = mc^2$ looks like potential 
internal energy, since the relativistic Lagrangian
\begin{equation}
L = -mc^2 \sqrt{1-\frac{v^2}{c^2}} - m G_N \frac{M}{r} =
-mc^2 \sqrt{1-\frac{v^2}{c^2}} - m\phi_{local},  
\end{equation}
may be approximately represented as follows:
\begin{equation}
L_a = T - V = \frac{mv^2}{2} - mc^2 - m\phi_{local},  
\end{equation}
where $V=m(c^2 + \phi_{local})=m(U_{global} + \phi_{local})$. The question is: how the special relativity formula  $E = \frac{mc^2}{\sqrt{1-\frac{v^2}{c^2}}}$ anticipates the global gravitation potential $c^2$? Probably, because the paragon theory has such ability. For example Dirac's relativistic equation for electron predicts automatically spin degrees of freedom and positrons; Maxwell electrodynamics
gives different example of the similar kind of anticipation: it has Poincar\'e invariant form before Einstein's recognition of the physical sense of this invariance. 

So, we will assume that just fundamental constant 
$c^2$ is defined by self-consistent potential of 
the whole Universe, not a mass in spite of Mach-Einstein 
assumption \cite{Le3}. This global 
potential has presumably a microscopic exhibition
(`tension' of the vacuum) too.
{\it Now we will assume that universal character of the
potential $U_{global} = c^2$ connected not
only with the space-time geometry, but with
the geometry of quantum state space. One reasonable
hypothesis is that the 
constant sectional curvature of $CP(N-1)$ should contain 
this world constant}. This assumption
leads to far going consequence: 
fundamental interactions presumably determined by 
dimensionless holomorphic sectional
curvature ${\cal{K}}$ of the projective Hilbert space 
$CP(N-1)$, using as the carrier of self-consistent 
vacuum states, should
be proportional to the fine structure constant 
${\cal{K}} = a \alpha = a \frac{e^2}{\hbar c}$
(see, for a comparison, the arguments by \cite{Heslot,CMP}
for the $\hbar$ introducing in the sectional
curvature of $CP(N-1)$). 
The universal and small dimensionless parameter
${\cal{K}}$ will be used in the next work
for the unified interaction problem.

Einstein emphasized that it is impossible to 
distinguish the ``kinetic energy'' from the 
potential energy that associated with $mc^2$ in the formula 
$E = \frac{mc^2}{\sqrt{1-\frac{v^2}{c^2}}}$.
It is because Einstein dealt with
closed system of light source and electromagnetic
wave packets in his thought experiment. The global
potential makes any system open, therefore
interpretation $mc^2$ as a potential energy
is possible. In quantum theory due to the vacuum
fluctuations, any system is open and isolated quantum
object is an important approximation.
It is interesting to find a pure quantum analog of
the formula for energy of some quantum model so,
that it is possible to distinguish kinetic energy and 
potential only approximately as well as 
in the classical formula for the material point. 
It may be the energy of soliton solution for
a non-liner wave equation. So-called `chiral models' 
give us the examples of such kind. One of the well known
`chiral model' is $CP(N-1)$ model.
\vskip .2cm
\section{Underlying Field CP(N-1) Structure}
\vskip .2cm
Super-relativity principles tells
that there exist some fundamental invariants (like
electric charge) concerning internal state space
geometry that do not depend on embedding of a
quantum system into background field. In fact this
principle has been established in particular cases 
in so-called geometric phases works. Mainly these 
themes associated with works of Berry
\cite{Berry1,Berry2}, Aharonov-Anandan \cite{AA1,AA2}, 
and Wilczek-Zee \cite{WZ}. In particular Aharonov and 
Anandan emphasized
there exist (infinite) class of equivalent Hamiltonians
(i.e. fields) ``that generate the motion  in 
$\cal{H}$ that project to the same close curve 
$\cal{C}$ in $\cal{P}$''. The geometric phase 
$\beta$ depends on the $\cal{C}$ only, i.e. 
it is invariant (Chern class) associated with 
connection and curvature in the fiber 
bundle. In fact only simplectic structure of the 
state space has been taken into account in these 
works. We will use the complex rays space $CP(N-1)$ like \cite{AA1,AA2}
instead of parameter space of Hamiltonian family 
\cite{Berry1,Berry2}. 
Berry \cite{Berry2} already pointed out that 
the symmetric part of the K\"ahler structure may 
have essential physical meaning. This is quantum 
geometric tensor defining Riemannian
structure on that manifold \cite{PV}.
I will discuss here only the case of
projective Hilbert space. 
 
The using $CP(\infty)$ or $CP(N-1)$ to 
the coherent states description is quite habitual.
In our interpretation $CP(N-1)$ is newly defined  
as a ``space of shapes'' \cite{SW} of the 
`elementary particles' (elementary FF excitations),
i.e. the space of representation of the coset manifold
$SU(N)/S[U(1) \times U(N-1)]$. 
It physically means that
there are some set of geometric invariants 
of the quantum state $|\psi> = \psi^i |i>$ 
with which it is possible to associate some `shape'
of quantum state under unitary dynamics. 
All conceivable states of the elementary vacuum
excitations in the Universe consist of `shapes' cum location
in space-time. Location in space-time is represented 
by the local variations of the global potential 
$<V|U_{global}|V> = c^2$.
In accordance with our assumption the unitary
super-multiplete $\Psi^k$ is capable to represent any
quantum particle in any background field. 
Therefore, there is the omnipresence
of the local unitary transformations 
$U_{(AB)} \in SU(N-1)$ capable
to transform particle A into particle B:
$|\Psi (B)> = U_{(AB)} |\Psi (A)> $.
This give us a possibility looking for
pre-dynamics in the compact manifold $CP(N-1)$ 
of $SU(N)$ group representation 
instead of space-time itself.    
We will looking for only relative amplitudes
$\pi^k = \frac{\psi^k}{\psi^j}$ presenting
points of $CP(N-1)$ and tangent fiber bundle
over them. Variations of points in $CP(N-1)$
correspond to local variations of the global potential
(initial conditions). 
The measure of the distance between two
coherent states in $CP(N-1)$  may be express 
through the quantum metric tensor
\begin{equation}
G_{ik^*} = {\cal{K}}^{-1} \frac{(1+\sum |\pi^s|^2) \delta_{ik}-\pi^{i^*} \pi^k}
{(1+\sum |\pi^s|^2)^2}.
\label{FS}  
\end{equation}

Now we should generalize Pancharatnam problem of the 
comparison of two polarized light beams. Namely, how
one can say that two quantum system are `in phase',
i.e. coherent? First Einstein
in his works in ideal quantum gas predicting Bose-
Einstein condensation realized that the conditions
of the constructive interference of de Broglie
waves in the ideal 
gas, are identical masses of atoms and their velocities \cite{Einst1,Einst2}. 
The problem of comparison of identical 
quantum particles (their dynamical variables 
in the different background fields) leads to the 
problem of the parallel transport of tangent vector
fields corresponding to different quantum dynamics.
The affine connection $\Gamma^i_{mn}$
defined as follows 
\begin{eqnarray}
\Gamma^i_{mn} = \frac{1}{2}G^{ip^*}
(\frac{\partial G_{mp^*}}{\partial \pi^n} +
\frac{\partial G_{p^*n}}{\partial \pi^m})  \cr
= - \frac{\delta^i_m \pi^{n^*} + \delta^i_n \pi^{m^*}}{1+\sum |\pi^s|^2},
\label{Gamma}  
\end{eqnarray}
where
\begin{equation}
G^{ik^*} = {\cal{K}}(1+\sum |\pi^s|^2)( \delta^{ik}+\pi^{i} \pi^{k^*})
\label{aFS}  
\end{equation}
for the parallel transport agreed with Fubini-Study
metric should be taken into account. 
The Riemannian curvature
\begin{eqnarray}
R^i_{klm^*} = - \frac{\partial \Gamma^i_{kl}}{\partial \pi^{m^*}} \cr =  \{\frac{\delta^i_k \delta^l_m + \delta^i_l \delta^k_m}
{1+\sum |\pi^s|^2}
-\pi^m \frac{\delta^i_k \pi^{l^*} + \delta^i_l \pi^{k^*}}{(1+\sum |\pi^s|^2)^2}\}
\label{Riemann}  
\end{eqnarray}
will be associated with the Yang-Mills fields
of the new kind defined on $CP(N-1)$.
\vskip .2cm
\section{Quantum Pre-Dynamics of the Fundamental Field}
\vskip .2cm
\subsection{Generalized Coherent States and Gauge Principle}
In the absence (at this stage) of space-time coordinates 
one need some `hidden' quantum pre-dynamics. However,
dynamics should be exhibited in a sub-manifold 
of quantum states especially identifying with ordinary space-time.
Due to the nonlinear character of $CP(N-1)$ (curvature)
we have self-interacting quanta of the vacuum excitations.
The gauge principle is the simplest way to establish
field equations corresponding to such interaction.
Initially it will be equation in the state space
and only after special `reduction' one could obtain
space-time representation of the field equations.

Pursuing to realize the `strong relativity to embedding fields'
principle, one should admit a wide class of the
functional frames capable to represent such fields.
Furthermore, dynamical variables should be covariant
operators relative functional frame variations.
That is one have to have a covariant ``functional
calculus''  \cite{Asht}. We will use such covariant
calculus in the projective Hilbert state space $CP(N-1)$
relative Fubini-Study metric. The discussion of the 
convergence problem in the case $CP(\infty)$ will 
be postponed for the future work.

Locally (in the state space) the gauge transformation 
mechanism may be described by 
the connection form in $CP(N-1)$
\begin{eqnarray}
\Omega^i_k= \Gamma^i_{km} \delta \pi^m. 
\label{omega} 
\end{eqnarray}
Different choice of the independent field variables 
$\pi^{\tilde{i}}$
(holomorphic diffeomorphism of a bounded domain 
$D \subset C^N$ onto a bounded domain $D' \subset C^N$, 
say transition for a different vacuum $\psi^k \neq 0$), 
\begin{eqnarray}
\pi^{\tilde{i}}= \pi^{\tilde{i}}(\pi^1,..., \pi^{N-1}),
\label{transpigen} 
\end{eqnarray}
such that
\begin{equation}
\frac{\partial \pi^{\tilde{i}}}{\partial \pi^{m*}} = 0
\end{equation}
corresponding to the variation of representation 
(variation of the functional reference frame), entails the 
non-Abelian gauge transformations law of the connection form $\Omega^{m}_n$
of the well known kind:
\begin{eqnarray}
\Omega^{\tilde{i}}_k=  J^{-1\tilde{i}}_m \Omega^{m}_n J^n_k +
J^{-1\tilde{i}}_t \delta J^t_k,
\label{transOm} 
\end{eqnarray}
where
\begin{eqnarray}
J^{\tilde{i}}_{m}  = \frac{\partial \pi^{\tilde i}}{\partial \pi^m} 
\label{Jacobian} 
\end{eqnarray}
is the matrix with holomorphic Jacobian . Transformations
of the connection form should be accompanied 
by the transformations of tangent vectors fields
{\it taking the place of the state vectors in the
tangent spaces intrinsically 
defined by the dynamical variables in our
interpretation}  
\begin{eqnarray}
T^{\tilde{i}}= J^{\tilde{i}}_m T^m = \frac{\partial \pi^{\tilde{i}}}{\partial \pi^m} T^m.
\label{tang} 
\end{eqnarray}
Hence, in general, the gauge transformations belongs to $GL(N-1,C)$ but in physical applications this group 
should be reduced to the $U(N-1)$. In particular,
an important example of such transformations applicable 
to the $\Phi_{\sigma}^i$ (see below). 
\begin{eqnarray}
\Phi_{\sigma}^{\tilde{i}}= J^{\tilde{i}}_m \Phi_{\sigma}^m = \frac{\partial \pi^{\tilde{i}} }{\partial \pi^m} \Phi_{\sigma}^m.
\label{tangent} 
\end{eqnarray}
I would like to emphasize here that non-Abelian gauge
transformation \ref{transOm} only looks 
like Wilczek-Zee gauge potential \cite{WZ} but 
is not identical to this one. Our non-Abelian
potential has geometric origin in cotangent space
$T^*_{\pi} CP(N-1)$ (affine connection form),
whereas Wilzcek-Zee potential 
originated by the instant Hamiltonian with
degenerated spectrum in Hilbert space $C^N$. 
Jacobian factors $ J^{\tilde{n}}_k $ are akin here to 
Shapere-Wilczek gauge transformations in
the space of shapes of deformable bodies \cite{SW}.  
However, one has simpler situation here 
since we deal not with arbitrary deformations of 
the swimming body but with the so-called
`generalized solid body' motion
with deformations of the ellipsoid of polarization
restricted by the coset structure 
$CP(N-1) = SU(N)/S[U(1) \times U(N-1)]$ of the dynamical
group $SU(N)$ \cite{Le1,Le2,Le3,Le4}.
The tangent vectors to $CP(N-1)$ determinate
the velocities of entanglement of infinitesimally
close amplitudes i.e. variations of vacuum state. 
For example the infinitesimal shifts generated by the one-
parameter groups in $1 \leq \sigma \leq N^2-1$ directions of $AlgSU(N)$ defined by coefficient functions as follows:
\begin{equation}
\Phi_{\sigma}^i = \lim_{\epsilon \to 0} \epsilon^{-1}
\biggl\{\frac{[\exp(i\epsilon \lambda_{\sigma})]_m^i \Psi^m}{[\exp(i \epsilon
\lambda_{\sigma})]_m^j
\Psi^m }-\frac{\Psi^i}{\Psi^j} \biggr\}=
\lim_{\epsilon \to 0} \epsilon^{-1} \{ \pi^i(\epsilon \lambda_{\sigma}) -\pi^i \}.
\end{equation}
Dynamical variables 
$D_\sigma =\Phi^i_{\sigma}\frac{\partial}{\partial \pi^i} + c.c.$ play role of quantum states in the local tangent space.
{\it We will distinguish two kinds of quantum states}:
`coherent quantum state' defined by $\pi^i$ local
coordinates of FF in $CP(N-1)$ and `dynamical quantum state'
in tangent space defined by the linear combinations of $SU(N)$ generators $\Psi(\pi) = a^{\sigma} D_{\sigma}$ in local coordinates. 
Thereby in the framework of the local state-dependent 
approach one can formulate a quantum
scheme with help more flexible mathematical structure 
than  matrix formalism. It means that matrix elements of 
transitions between {\it two arbitrary far states} are 
associated with, in fact, bi-local dynamical 
variables that
bring a lot of technical problems in quantum field area.
(See notes in this spirit in \cite{PV}).
However the infinitesimal local dynamical 
variables related to deformations of quantum
states are well  defined in projective Hilbert 
space as well as quantum states itself. 
They are local tangent vector fields to the 
projective Hilbert space $CP(N-1)$ which 
correspond to the group variation of the 
relative `Fourier components', i.e. 
{\it generators (differential operators 
of first order)} \cite{Le1,Le2,Le3}. 
In the local coordinates $\pi^i = \frac{\Psi^i}{\Psi^0}$
one can build the infinitesimal generators of the
Lie algebra $AlgSU(N)$.
Then one has to use explicit form 
$\Phi^i_\sigma$ for $N^2-1$ of infinitesimal generators of 
the Lie algebra $AlgSU(N)$. For example for the three-level
system, algebra $SU(3)$ has 8 infinitesimal generators which 
are given by the vector fields:
\begin{eqnarray}
D_1(\lambda)=i \frac{\hbar}{2}[[1-(\pi^1)^2]\frac{\delta}{\delta \pi^1}
-\pi^1 \pi^2 \frac{\delta}{\delta \pi^2}
-[1-(\pi^{1*})^2]\frac{\delta}{\delta \pi^{1*}}
+\pi^{1*} \pi^{2*} \frac{\delta}{\delta \pi^{2*}}] ,  \cr
D_2(\lambda)=- \frac{\hbar}{2}[[1+(\pi^1)^2]\frac{\delta}{\delta \pi^1}
+\pi^1 \pi^2 \frac{\delta}{\delta \pi^2}
+[1+(\pi^{1*})^2]\frac{\delta}{\delta \pi^{1*}}
+\pi^{1*} \pi^{2*} \frac{\delta}{\delta \pi^{2*}}] , \cr
D_3(\lambda)=-i \hbar[\pi^1 \frac{\delta}{\delta \pi^1}+\frac{1}{2}\pi^2
\frac{\delta}{\delta \pi^2}
+\pi^{1*} \frac{\delta}{\delta \pi^{1*}}+\frac{1}{2}\pi^{2*} \frac{\delta}{\delta
\pi^{2*}}], \cr
D_4(\lambda)= i\frac{\hbar}{2}[[1-(\pi^2)^2]\frac{\delta}{\delta \pi^2}
-\pi^1 \pi^2 \frac{\delta}{\delta \pi^1}
-[1-(\pi^{2*})^2]\frac{\delta}{\delta \pi^{2*}}
+\pi^{1*} \pi^{2*} \frac{\delta}{\delta \pi^{1*}}] , \cr
D_5(\lambda)= -\frac{\hbar}{2}[[1+(\pi^2)^2]\frac{\delta}{\delta \pi^2}
+\pi^1 \pi^2 \frac{\delta}{\delta \pi^1}
+[1+(\pi^{2*})^2]\frac{\delta}{\delta \pi^{2*}}
+\pi^{1*} \pi^{2*} \frac{\delta}{\delta \pi^{1*}}], \cr
D_6(\lambda)=i\frac{\hbar}{2}[\pi^2 \frac{\delta}{\delta \pi^1}
+\pi^1 \frac{\delta}{\delta \pi^2}
-\pi^{2*}\frac{\delta}{\delta \pi^{1*}}
-\pi^{1*} \frac{\delta}{\delta \pi^{2*}}] , \cr
D_7(\lambda)=\frac{\hbar}{2}[\pi^2 \frac{\delta}{\delta \pi^1}
-\pi^1 \frac{\delta}{\delta \pi^2}
+\pi^{2*}\frac{\delta}{\delta \pi^{1*}}
-\pi^{1*} \frac{\delta}{\delta \pi^{2*}}] , \cr
D_8(\lambda)=-\frac{3^{1/2}}{2}i\hbar[\pi^2 \frac{\delta}{\delta \pi^2}
-\pi^{2*} \frac{\delta}{\delta \pi^{2*}}]. 
\label{D8} 
\end{eqnarray}
In general, such dynamical variables 
$\Phi^i_\alpha, \Phi^i_\beta $ define the curvature 
in 2-dimension direction $(\alpha,\beta)$ 
\begin{eqnarray}
R(D_\alpha,D_\beta)X^k = 
[\nabla_{D_\alpha},\nabla_{D_\beta}] X^k -
\nabla_{[D_\alpha,D_\beta]} X^k \cr
= \{(D_\alpha \Phi^i_\beta - D_\beta \Phi^i_\alpha )
\Gamma^k_{in}  
-(\Phi^i_\beta \Phi^{*s}_\alpha -
\Phi^i_\alpha \Phi^{*s}_\beta)\frac{\partial \Gamma^k_{in}}{\partial \pi^{*s}} \cr +
\Phi^m_\alpha \Gamma ^k_{mp}\Phi^i_\beta \Gamma ^p_{in}-
\Phi^i_\beta \Gamma ^k_{ip}\Phi^m_\alpha \Gamma ^p_{mn}
- C^\gamma_{\alpha \beta} \Phi^i_\gamma \Gamma^k_{in}\}X^n 
\cr =
\{(D_\alpha \Phi^i_\beta - D_\beta \Phi^i_\alpha )
\Gamma^k_{in}  
-(\Phi^i_\beta \Phi^{*s}_\alpha -
\Phi^i_\alpha \Phi^{*s}_\beta) R^k_{i*sn} \cr +
\Phi^m_\alpha \Gamma ^k_{mp}\Phi^i_\beta \Gamma ^p_{in}-
\Phi^i_\beta \Gamma ^k_{ip}\Phi^m_\alpha \Gamma ^p_{mn}
- C^\gamma_{\alpha \beta} \Phi^i_\gamma \Gamma^k_{in}\}X^n. 
\label{R}  
\end{eqnarray}
Here $C^{\gamma}_{\alpha \beta}$ are of the $SU(N)$ group 
structure constants \cite{Le4}. 
Now we can introduce the follows expression 
\begin{eqnarray}
F_{\alpha \beta} = R (D_\alpha, D_\beta) X^k 
\frac{\partial}{\partial \pi^k}  
+ c.c.,
\label{ym}
\end{eqnarray}
which is the analog of Yang-Mills fields (with the 
unknown $X^k$) of the gauge 
potential associated with the intrinsic affine 
$CP(N-1)$ connection 
\begin{equation}
\Omega^i_k= \Gamma^i_{km}\delta \pi^m = 
\frac{1}{\hbar} \Gamma^i_{km} \Phi_{\sigma}^m(\pi) 
\delta a^{\sigma}, 
\label{Omega} 
\end{equation}
where $\delta a^{\sigma}$ is a variation of the action. 
In comparison with Berry
singular potential $A_n(R)=<n(R)|\nabla_R n(R)>$
\cite{Aitch} this potential is not 
singular and the source of non-Abelian field is 
the curvature of $CP(N-1)$.
This non-Abelian field is expressed in local 
`slow' coordinates $\pi^i$ as well as
Berry ``monopole'' field in terms of $R$-
coordinates. I treat these functional fields 
as a pre-dynamical, initial condition variations,
or tunneling processes, in spite of the 
ordinary gauge fields constructions where fields
are functions of the space-time coordinates.

So, the super-relativity principle formulated
as the covariance of the field theory relative 
holomorphic
local coordinates $\pi^i$ variations is realized   
by the state-dependent gauge principle concerning
initially interacting quantum objects \cite{Jones}.
But in our case we have much more general kind of the
non-Abelian gauge transformations 
\ref{transpigen}.
Here one has the literally covariant derivatives
associated with Fubini-Study metric \ref{FS} 
in spite of
the commonly used generalized gauge theory 
notions. The hermitian hamiltonian defined by dynamical 
variables (tangent vectors) 
$T^p(\pi)$ may then be expressed as follows:
\begin{eqnarray}
H^{ps^*} =  G^{ik^*} (\frac{\partial
T^p}{\partial \pi^i} + \Gamma^p_{in} T^n)
( \frac{\partial
T^{s*}}{\partial \pi^{k*}}+ 
\Gamma^{s*}_{k*q*} T^{q*}) = \cr
G^{ik^*} T^p_{;i} T^{s*}_{;k*}.
\end{eqnarray}
It is similar to the scalar composition 
\begin{equation}
{\cal{H}} =  G^{ik^*} \frac{\partial U}
{\partial \pi^i} \frac{\partial U}
{\partial \pi^{k*}}
\end{equation}
generalizing Weinberg's multiplication $a*b$ in 
his nonliner 
modification of the quantum mechanics \cite{W}. 
{\it In fact, however, 
one has here the formulation of the covariant 
variational problem for the functional vector field
instead of `scalar'-functional of the traditional
variational calculus}. 
So, hamiltonian created by the tangent vector field 
in the $\sigma$ - direction of iso-space of adjoint
representation looks like
\begin{eqnarray}
H^{ps^*}_{\sigma} = G^{ik^*} \Phi^p_{\sigma ;i}
\Phi^{s*}_{\sigma ;k*}.
\end{eqnarray}
Then in the ``b-direction'' \cite{Le1,Le3}, 
i.e. along the direction of the geodesic flow 
generators, one has
\begin{eqnarray}
H^{ps^*}_b = G^{ik^*} \Phi^p_{b;i}
\Phi^{s*}_{b;k*} = 0,
\end{eqnarray}
since these tangent vectors fields are parallel 
transported along geodesics. Since all geodesics
in $CP(N-1)$ issued from the original point $\pi^k_0$
are mutually transforming by the gauge group
$H=U(1) \times U(N-1)$, it is enough to proof
it for $CP(1)$. In the local coordinates $\pi^k$
equation for geodesics is as follows:
\begin{eqnarray}
\frac{d^2 \pi}{ds^2} - \frac{2 \pi^*}{1+|\pi|^2}
(\frac{d\pi}{ds})^2 = 0,
\end{eqnarray}
where $s$ is the length of a curve in $CP(1)$.
It has solution a $\pi(s) = \exp{i \alpha} \tan{s}$
with arbitrary constant $\alpha$.

Let's put $\alpha = 0$, then $\pi(s) = \tan{s}$
and $\frac{d\pi}{ds} = \frac{1}{\cos^2{s}}$.
On the other hand corresponding vector field
representing the generator `OY-rotation' is
$1+\pi^2 = 1+\tan^2{s} = \frac{1}{\cos^2{s}}$.
It is clear we have the coincidence (parallel transport) of
the generator of rotation (see two-level analog of 
\ref{D8} in \cite{Le1}) and the tangent vector field
anywhere along corresponding geodesic of $CP(1)$.
If one put $\alpha = i \pi/2$ it is easy to see
that the similar calculations give us  
$\frac{d\pi}{ds} = \frac{i}{\cos^2{s}}$ that
corresponds to the parallel transporting 
generator of the `OX-rotation':
$i(1-\pi^2)=i(1 - \exp{i \pi} \tan^2{s}) = \frac{i}{\cos^2{s}}$.
So, one has $2s = N-1$
complex vector fields generated zero-hamiltonian 
in the strong, i.e. in the non-average value sense.
Then one has in the $4s^2 = (N-1)^2 $ real  
``h-directions'' generated by the 
vector fields 
$D_h = \Phi^i_h \frac{\partial}{\partial \pi^i}+c.c. $. 
It gives the finite value hermitian matrix 
\begin{equation}
H^{ps^*}_h = G^{ik^*} \Phi^p_{h;i}
\Phi^{s*}_{h;k*} \neq 0,
\end{equation}
representing the class of the covariant local 
Hamiltonians in the tangent space. 
I would like to note for N-level system not only
dipole fields (like $H_x,H_y,H_z$) are actual but 
multipole components (like $Q_{xx}, Q_{xy}$), etc., too.
The origin of these multipole field components for spin
is anisotropy of magnetic materials. 
The eight gluons fields take the place
of these multipole `angle-field' components for the 
color quarks of QCD, etc. 
Hence, pulse acquired
by the quantum N-level system during an interaction 
process, has, in general, complicated 
dynamical structure \cite{Ostrov}. One can think even that
quantum particles themselves represent the {\it multipole
action pulse of the FF vacuum excitations}. 
Our target is to find some
dynamical covariant and invariant variables associated with
the geometry of the N-level coherent state space.
\vskip .2cm
\subsection{Quantum Mechanics As The Tangent Approximation}
\vskip .2cm 
Let us discuss now briefly construction of the liner
quantum systems of the ordinary quantum mechanics
in the tangent fiber bundle over $CP(N-1)$.
As before, we will discuss now only finite dimension
version of the theory, infinite dimension case will
be postponed to the special publication.

I will use for the simplicity the vacuum vector
in follows form: 
\begin{eqnarray}
|V> = \left( 
\matrix {R_{vac} \exp{(i\gamma)}  \cr
0 \cr
0 \cr
. \cr
. \cr
. \cr
0
}
\right).
\end{eqnarray}
Corresponding local coordinates in $CP(N-1)$ 
is simply
$(\pi^1,...,\pi^i,...,\pi^{N-1}) = (0,...,0)$. 
Any tangent vector corresponding some
dynamical variable (a self vector of the
hernitian Hamiltonian, for example), we take 
in the form
\begin{eqnarray}
|\xi> = \left( 
\matrix {0 \cr
\xi^1 \cr
\xi^2 \cr
. \cr
. \cr
. \cr
\xi^{N-1}
}
\right).
\end{eqnarray}
It is clear that they are orthogonal relative the standard
scalar product in `surrounding' Hilbert space $C^N$:
$<V|\xi> = 0$. In the standard perturbation theory
$|V>$ corresponds to the unperturbed solution, and
$|\xi>$ corresponds to one of the approximation
solutions orthogonal to the vacuum. 
If one looks on $|\xi>$ as a finite
variation of the vacuum (`tangent extrapolation' 
of the infinitesimal generator), than
the sum is the state vector
\begin{eqnarray}
|\phi> = |V> + |\xi> = \left( 
\matrix {R_{vac} \exp{(i\gamma)}  \cr
\xi^1 \cr
\xi^2 \cr
. \cr
. \cr
. \cr
\xi^{N-1}
}
\right).
\end{eqnarray}
For us will be interesting not this `imaginary'
state vector, but its projection on $CP(N-1)$.
In order to do it one has to normalize $|\phi>$
to the `vacuum radius' $R_{vac}$. It is easy to see
that coordinates is as follows:
\begin{eqnarray}
|\phi_{\cal{N}}> = \cal{N}\left( 
\matrix {R_{vac}\exp{(i\gamma)}  \cr
\xi^1 \cr
\xi^2 \cr
. \cr
. \cr
. \cr
\xi^{N-1}
}
\right),
\end{eqnarray}
where ${\cal{N}}=\frac{R_{vac}}{\sqrt{R_{vac}^2+\sum{|\xi^s|^2}}}$.
The local coordinates of $|\phi_{\cal{N}}>$ in the
chart originated at the new ``physical vacuum''
one will find
$\pi^{'i} = \frac{\xi^i \exp{(-i\gamma)}}{R_{vac}}$. 
In fact the `tangent extrapolation' is illegal 
since an infinitesimal operator (`Hamiltonian of
perturbation') does not define the finite 
unitary (norm-preserving) variation
of the vacuum vector. Only infinite set of the 
infinitesimal, continuous,      
successive unitary transformations lead to the
state with the local coordinates $\pi^{'i}$.
Furthermore, there are continuum pathes in
the Hilbert projective space connecting the 
vacuum and the perturbed state.
In particular, one can chose the geodesic in $CP(N-1)$
connecting the point 
$(\pi^1,...,\pi^i,...,\pi^{N-1}) = (0,...,0)$
with the point 
$\pi^{'i} = \frac{\xi^i \exp{(-i\gamma)}}{R_{vac}}$
\cite{Le1,Le3}.
Probably the geodesic motion in projective
Hilbert space $CP(N-1)$ physically corresponds
to the tunneling process going in a `tunneling time'.
It may demolish the necessity in wormholes in 
space-time since quantum tunneling goes in
the state space acquiring thereby physical reality. 

Therefore, a realistic physical problem with
interacting terms should be formulated in the
spirit of the sketch given above. In particular,
dynamical quantum state creeps along geodesic
of $CP(N-1)$ from one vacuum to another and
will belongs to the different tangent Hilbert 
spaces at each instant of the `tunneling time'
\cite{Le5}. 

It is absolutely clear that any
quantum setup for generating or registration
of the quantum objects is built from physical
fields and, hence, its coordinates correspond
to some point of $CP(N-1)$. What we see in the
space-time (say, cloud chamber) is merely 
very poor picture. Deep processes going in the
state space. It looks like any quantum object is 
placed in the
self-consistent global (cosmic) potential 
$<V|U_{global}|V> = c^2$ and all attempts
to disturb this equilibrium state leads to
the `reaction' which manifestation is the inertia
of the quantum object.  
Reaction of the quantum setup 
on the some particle emission is an example
of such kind. New coordinates will correspond
after the emission process to the particle and 
to the setup too. One may assume that there is   
{\it `inertia' of the quantum state and its elasticity 
under deformation}. This elasticity presumably 
determines the inertia (mass) of the quantum 
system \cite{Le3}.
The structure of this
state has some quantum integrals of motion 
(`charges') and, hence, 
should be invariant under embedding into an external
field.
\vskip .2cm
\subsection{Topological and Tunneling Integrals}
\vskip .2cm 
There are two kinds of the physically
interesting integrals arising over
base manifold of the tangent fiber bundle $CP(N-1)$.
The first kind is the characteristic classes by Chern
of the coholonomy groups $c_k \in H^{2k}(CP(N-1))$.  
Corresponding topological invariants are 
integrals 
\begin{equation}
I_k = \int_{M_{2k} \subset CP(N-1)} \tilde{c}_k 
\end{equation}
from the forms
\begin{eqnarray}
\tilde{c}_k = \frac{1}{2\pi i} Tr(R(D_{\alpha_1}, D_{\beta_1}) \wedge R(D_{\alpha_2}, D_{\beta_2})\wedge ...
\wedge R(D_{\alpha_k}, D_{\beta_k})) 
\end{eqnarray}
representing a new kind of geometric phases.
The formal `Yang-Mills equations' obtained as 
a requirement of the elimination of the covariant
derivative from the analog of Yang-Mills field \ref{ym}
\begin{eqnarray}
\nabla_{\beta} R(D_{\alpha}, D_{\beta})X^k = 0 
\end{eqnarray}
should be analyzed carefully in the future 
investigations from the physical point of view.    

The second kind is the `tunneling integrals', i.e.
invariants connected with Fubini-Study metric
\ref{FS}. Any scalar product 
$(\theta(\pi)|\xi(\pi)) =  G_{ik*}(\pi) \theta^i(\pi) \xi^{k*}(\pi)$, 
where $|\xi(\pi)), |\theta(\pi)) $ belong to the 
tangent space at the origin point, is invariant 
relative the parallel transport in $CP(N-1)$. Besides
such kind of the local invariants should exist  
global invariants due to the global subgroup
$H = U(1) \times U(N-1)$ of the gauge symmetry 
on the $CP(N-1)$. This group transforms one
geodesic to another, therefore states belonging
to these geodesics should have some invariant
characteristics.  
One can associate that state characteristics 
with the shape of the ``ellipsoid of polarization''. 
The full set of the ellipsoid parameters comprises 
of state vector $4s$ real variables itself
and the $4s^2$ real parameters of ``orientation'' of 
the ellipsoid relative the quantum frame.
Here we derive corresponding formulas for such ellipsoid.
First of all let us remind formulas for the simplest case
of the two level system $N = 2; N = 2s+1$.  
Let us discuss some two level quantum system in the basis
$|1>, |2>$ with the coefficients 
$\psi^1=|\psi^1| e^{i \alpha_1}=|\psi^1| e^{i \alpha'_1 + \tau}$,
$\psi^2=|\psi^2| e^{i \alpha_2}=|\psi^1| e^{i \alpha'_2 + \tau}$. If one defines $\rho^k=Re \psi^k$ then it is easy to verify
that follows equation of ellipse takes the place:
\begin{equation}
(\frac{\rho^1}{|\psi^1|})^2 + (\frac{\rho^2}{|\psi^2|})^2 
- 2 \frac{\rho^1}{|\psi^1|} \frac{\rho^2}{|\psi^2|} cos \delta 
= sin^2 \delta,
\end{equation}
where $\delta = \alpha'_2 - \alpha'_1$.
This ellipse (its shape) is invariant relative isotropy group
of the spin $s=1/2$ ground state (vacuum) 
\begin{eqnarray}
|1> = \left( 
\matrix {1 \cr
0
}
\right).
\end{eqnarray}
Our target now is to get general invariant for N-level system.
This problem was risen by V.Ostrovskii in the private  
talk in the connection with his interesting article 
\cite{Ostrov}. I think that it is natural to build
the follows quadric formula:
\begin{eqnarray}
\sum_{i=1}^N (\frac{\rho^i}{|\psi^i|})^2 - \frac{2}{N-1}
\sum_{i<j}^{(N^2-N)/2} \frac{\rho^i}{|\psi^i|}\frac{\rho^j}{|\psi^j|}\cos{(\delta_{ij})}
=\frac{1}{N-1}\sum_{i<j}^{(N^2-N)/2} \sin^2{(\delta_{ij})}
\end{eqnarray}
where $\delta_{ij} = \alpha_i - \alpha_j$.
I have found that it is ellipsoid, but I could not
to prove that its shape is really invariant relative
the isotropy group $H = U(1) \times U(N-1)$
of the ground state
\begin{eqnarray}
|1'> = \left( 
\matrix { 1  \cr
0 \cr
0 \cr
. \cr
. \cr
. \cr
0
}
\right).
\end{eqnarray}
It is important to verify
the invariance fact of the ellipsoid shape. I will
sincerely grateful for somebody who will solve this problem. 
If my assumption is correct, it lead to the mechanism
of the reconstruction of the unitary symmetry $SU(N)$. 
Such reconstruction of the unitary symmetry represents the concrete mechanism of symmetry breakdown by the coset 
transformations $G/H=SU(N+1)/S[U(1)\times U(N)]$
up to isotropy  group  $H=U(1)\times U(N)$. Therefore the shape
of the ellipsoid of polarization will a new integral invariant
of unitary quantum dynamics. Then will be possible
to try to connect this mechanism of the unitary 
symmetry breakdown with the problem of mass split 
effects in the unitary 
fundamental multiplet of ``elementary particles''.

\vskip .2cm
\subsection{Quantization in CP(N-1)}
\vskip .2cm
Let us use the local projective representation 
of $SU(N)$ generators acting upon the smooth (holomorphic) functions of the coherent state instead of matrices acting upon
N-component original state vector $|\psi>$ of FF. These $D_{\sigma}$ generators are the tangent vector fields
to the $CP(N-1)$ \cite{Le1,Le4,Le5}.
Hereby 
the boson `second quantization' over $CP(N-1)$ 
takes the place quite naturally.  But one should take 
into account the ambiguity of quantization axes notion for $SU(N)$.
Since the rank of $AlgSU(N)$ is $r = N-1 = 2S$ and
number of the local independent charts for the covering
$SU(N)$ is $l = r + 1 = N = 2S + 1$. Single axes of
quantization there is only for a two level system 
($S = 1/2$). 
The choice of the chart marks the 
`open channel' ($\psi^j \neq 0$)  in a `filter' experiment,
and it determines a local `vacuum' of our model.
On the pre-dynamics level one has not a possibility
to measure space-time or energy-momentum values.
It means that there exist only 
{\it the operator of action} 
\begin{equation}
A=\hbar \pi^{*i}\frac{\partial}{\partial \pi^{*i}},
\label{action}
\end{equation}
which creates the 
quanta of action. Then the character of the distribution
of these action quanta among degrees of freedom
will be realized as specific quantum particles.

Commutation relations for ``geometrical bosons'' 
is as follows:
\begin{equation}
[\frac{\partial}{\partial \pi^k},\pi^i]_-=\delta^i_k.
\label{comm}
\end{equation}
The standard Fock representation is achievable under
the definition of vacuum state by a holomorphic 
function $F_{vac}$. Over compact connected manifold
like $CP(N-1)$ such function is a constant.
Then, since $\frac{\partial F_{vac}}{\partial \pi^{*k}}=0$
one can introduce the function of excitations
of different degrees of freedom 
$F(s_1,...,s_{\cal{N}})=\pi^{*s_1} \pi^{*s_2}...
\pi^{*s_{\cal{N}}}F_{vac}$ and the function of a 
multifold excited degree of freedom
$F(s;\cal{N})$ $=(\pi^{*s})^{\cal{N}}F_{vac}$.
For this function one has the equidistant
spectrum of {\it action}:
\begin{eqnarray}
A F(s;0)=0; \cr 
A F(s;1)=\hbar F(s;1); \cr
A F(s;2)=2\hbar F(s;2); \cr
. \cr
. \cr
. \cr
A F(s;{\cal{N}})
={\cal{N}} \hbar F(s;{\cal{N}}).
\label{oscill}
\end{eqnarray}

\end{document}